\documentclass[conference]{IEEEtran}

\IEEEoverridecommandlockouts

\usepackage{cite}
\usepackage{amsmath,amssymb,amsfonts, amsthm}
\usepackage{algorithmic}
\usepackage{graphicx}
\usepackage{textcomp}
\usepackage[utf8]{inputenc}
\usepackage[T1]{fontenc}
\usepackage{multicol}
\usepackage{multirow}

\def\BibTeX{{\rm B\kern-.05em{\sc i\kern-.025em b}\kern-.08em
    T\kern-.1667em\lower.7ex\hbox{E}\kern-.125emX}}
    
\begin{document}

\title{High-Quality Disjoint and Overlapping Community Structure in Large-Scale Complex Networks}

\author{\IEEEauthorblockN{1\textsuperscript{st} Eduar Castrillo}
\IEEEauthorblockA{\textit{Departamento de Ingeniería de Sistemas} \\
\textit{Universidad Nacional de Colombia}\\
Bogotá D.C., Colombia \\
emcastrillov@unal.edu.co}
\and
\IEEEauthorblockN{2\textsuperscript{st} Elizabeth León}
\IEEEauthorblockA{\textit{Departamento de Ingeniería de Sistemas} \\
\textit{Universidad Nacional de Colombia}\\
Bogotá D.C., Colombia \\
eleonguz@unal.edu.co}
\and
\IEEEauthorblockN{3\textsuperscript{st} Jonatan Gómez}
\IEEEauthorblockA{\textit{Departamento de Ingeniería de Sistemas} \\
\textit{Universidad Nacional de Colombia}\\
Bogotá D.C., Colombia \\
jgomezpe@unal.edu.co}
}

\maketitle

\begin{abstract}
In this paper, we propose an improved version of an  agglomerative hierarchical clustering algorithm that performs disjoint community detection in large-scale complex networks. The improved algorithm is achieved after replacing the local structural similarity used in the original algorithm, with the recently proposed Dynamic Structural Similarity. Additionally, the improved algorithm is extended to detect fuzzy and crisp overlapping community structure. The extended algorithm leverages the disjoint community structure generated by itself and the dynamic structural similarity measures, to compute a proposed membership probability function that defines the fuzzy communities. Moreover, an experimental evaluation is performed on reference benchmark graphs in order to compare the proposed algorithms with the state-of-the-art.
\end{abstract}


\section{Introduction}

Networks are ubiquitous because they conform the backbones of many complex systems, such like social networks, protein-protein interactions networks, the physical Internet, the World Wide Web, among others \cite{erciyes2014complex}. In fact, a complex system can be modeled by a complex network, in such a way that the complex network represents an abstract model of the structure and interactions of the elements in the complex system \cite{newman2011structure}. For example, in social networks like Facebook, the network of user-user friendships, the social groups and the user reactions to posts, can be modeled as complex networks.

One important property of complex networks is the community structure. Detecting the community structure of a complex network is the task of grouping the set of nodes in such a way that nodes in the same group are more similar to each other than to those in the other groups. This particular task is known as \emph{Community Detection} and in this context, a group of nodes is denominated a \emph{Community}. Intuitively, a community can be defined as a group of nodes where there are more edges inside the community than edges towards the rest of the complex network \cite{erciyes2014complex}.

Community Detection is not a specific algorithm, but the generic task to be solved, that is why in the past two decades several algorithms for community detection have been developed. Usually, the community detection algorithms differ in their conception of community or how efficiently find them. Moreover, the community detection algorithms can be categorized according to their functionality as hierarchical clustering, graph partitioning, partitional clustering, spectral methods, optimization techniques, dynamic methods, among others \cite{fortunato2016community}.

With the arrival of the Big Data era a vast amount of  data is generated continuously. For example, many large-scale complex networks with thousands and millions of nodes and edges are generated from different sources such as social networks, peer-to-peer networks, the internet, etc \cite{leskovec2016snap, rossi2015network}. Algorithms are then required to handle data efficiently, so it is necessary to design and develop efficient methods to handle community detection on big networks, while keeping high quality results in terms of the detected community structure. However, algorithms for community detection with high quality results in small or medium sized networks are not suitable to work with large networks due to their high computational complexity  (e.g., Agglomerative Hierarchical Walktrap \cite{pons2005computing}, Divisive Hierarchical Girvan-Newman \cite{girvan2002community}, Information Theory Infomap \cite{rosvall2008maps}). On the other hand, efficient algorithms in terms of computational complexity, compute community structure with some limitations, for example, the Louvain method \cite{blondel2008fast} and its known problem of resolution limit \cite{fortunato2007resolution}, Label Propagation \cite{raghavan2007near} and its convergence problems due to the random nature of its functioning, or Attractor \cite{shao2014community} and the convergence problems presented in its underlying dynamic system, making it a slow algorithm in practice. Moreover, efficient algorithms that compute high quality community structure are not easy to tune in practice because they are too sensitive to the initial conditions given by the input parameters, for example, supervised algorithms that require the number of communities in advance as input parameter (this information is usually unknown in practice and is not easy to estimate) \cite{li2015uncovering, lofgren2016personalized}, or algorithms that strongly depend on parameters that are not easy to tune in practice \cite{xu2007scan, chang2017mathsf}. 

An algorithm that overcomes some of the aforementioned drawbacks is the HAMUHI algorithm \cite{castrillo2017fast}. HAMUHI is a novel fast heuristic algorithm for multi-scale hierarchical community detection that can be applied efficiently to large-scale complex networks. Compared to the state-of-the-art, this algorithm provides better trade-off among quality of the results, running time and easy to use. However, the quality of the resulting community structure is susceptible of being improved. 

In this paper, we propose an improved version of the HAMUHI Algorithm. The improved version of HAMUHI is achieved after replacing the local structural similarity used in it, with the recently proposed Dynamic Structural Similarity \cite{castrillo2018DSS}. Moreover, HAMUHI is further extended to leverage the disjoint community structure generated by itself to efficiently detect fuzzy and crisp overlapping community structure, while maintaining the same computational complexity.

The rest of this paper is organized as follows: In section \ref{sec_related_work} we briefly survey the related work about community detection. Section \ref{sec_hamuhi} presents in detail the proposed algorithms for disjoint and overlapping community detection. In Section \ref{sec_experiments} several experiments are performed in order to test the proposed algorithms and to compare them with the state-of-the-art. Finally, Section \ref{sec_conclusions} draws some conclusions about this research work.

\section{Related Work}
\label{sec_related_work}

\subsection{Graph Model}

\theoremstyle{definition}
\newtheorem{defn}{DEFINITION}

\begin{defn}
Let $G = (V, E)$ be a graph with set of vertices $V$, set of edges $(u,v) \in E$ such that $u, v \in V$. In the case of undirected graphs, the edges $(u, v)$ and $(v, u)$ are considered the same. For the rest of this paper we suppose $G$ is an undirected graph, unless other type of graph is explicitly mentioned.
\end{defn}

\begin{defn}
The structural neighborhood of a vertex $u$, denoted by $N(u)$, is defined as the open neighborhood of $u$; that is $N(u) = \{v \in V | (u, v) \in E\}$. Additionally, the closed structural neighborhood, denoted by $N[u]$, is defined as $N[u] = N(u) \cup \{u\}$.
\end{defn}

\begin{defn}
The degree of a vertex $u$, denoted by $d[u]$ and $d(u)$, is basically the cardinal of the structural neighborhood of $u$; that is $d[u] = |N[u]|$ and $d(u) = |N(u)|$.
\end{defn}

\subsection{Community Structure Model}

\begin{defn}
\label{def_community_structure}
The \emph{Community Structure} of a graph $G$, denoted by $C(G)$, is a set of communities extracted from $G$, i.e., $C(G) = \{C_i : C_i \subseteq V\}$. Moreover, each node must belong to at least one community, that is $\bigcup_{\substack{C_i \in C(G)}} C_i = V$. The community structure can be classified into two main categories: disjoint and overlapping community structure. 
\end{defn}

\begin{defn}
\label{def_disjoint_community}
$C(G)$ is a disjoint community structure if $\bigcap_{\substack{C_i \in C(G)}} C_i = \emptyset$, otherwise $C(G)$ is denominated an overlapping community structure. In other words, in the case of disjoint community structure, a node can belong to maximum one community, while in the overlapping community structure a node can belong to one or more communities.
\end{defn}

\begin{defn}
\label{def_fuzzy_community}
A \emph{Fuzzy Overlapping Community Structure} $C^f(G)$, is an overlapping community structure in which each community $C_i \in C^f(G)$ is a fuzzy set $(C_i, f_i)$, with membership function $f_i:C_i\mapsto[0, 1]$. For each node $u \in C_i$, the value $f_i(u)$ is called the probability of membership of $u$ in the community $C_i$.
\end{defn}

\begin{defn}
\label{def_crisp_community}
A \emph{Crisp Overlapping Community Structure} $C^\alpha(G)$, is an overlapping community structure in which each community $C_{i}^\alpha \in C^\alpha(G)$ is a crisp set obtained with an $\alpha$-cut of $(C_i, f_i)$, i.e., $C_{i}^\alpha = \{u \in C_i: f_i(u) \geq \alpha \}$.
\end{defn}

\subsection{State-of-the-art Algorithms}
\label{sota}

This research try to deal with unsupervised community detection in large-scale complex networks, for that reason, a special emphasis is made on the state-of-the-art unsupervised algorithms that can be efficiently applied in large-scale scenarios. An algorithm is considered unsupervised, if the number of communities to detect is not required as input parameter. Moreover, a particular algorithm can be considered efficient, if its time complexity is linear in terms of the number of edges in the input graph, i.e., time complexity of $O(|E|)$ in the worst or average case. 

Next, a general classification of the state-of-the-art algorithms for disjoint and overlapping community detection is briefly presented.

\textbf{Optimization Techniques:} The idea behind these methods is that a good community structure must present high values of the Modularity score \cite{newman2004finding}. Multilevel \cite{blondel2008fast} is a fast modularity optimization algorithm that performs an agglomerative hierarchical approach composed of two phases: network collapse and greedy optimization. These two algorithms find good local optima of the modularity. However, it has been proved that optimizing the modularity yields to the problem of resolution limit, making the modularity-based methods unable to detect communities smaller than a certain size that depends on the size of the network \cite{fortunato2007resolution}. Infomap \cite{rosvall2008maps} applies a greedy technique to minimize an objective function called the map equation. The map equation quantifies the information needed to represent a random walker in a network using a two-level nomenclature. Infomap can detect both, disjoint and overlapping community structure. Experimentally, Infomap has shown high computational complexity in large-scale complex networks \cite{castrillo2017fast}. Several multi-resolution modularity definitions have been proposed, but they tend to divide big communities to favor the small communities, that means, the multi-resolution modularity definitions also present resolution limit \cite{fortunato2016community}. In fact, any algorithm that optimizes partition quality functions, like modularity, will yield to some resolution limit \cite{fortunato2007resolution, Kawamoto2015EstimatingTR}.

\textbf{Information Propagation:} Label Propagation proposed by Raghavan et al. \cite{raghavan2007near} simulates the diffusion of information (labels) through the network. At the beginning, each node is labeled with a unique value, then iteratively each node takes the most frequent label in its neighborhood and the process continues until convergence. The results provided by Label Propagation are sometimes unpredictable and the whole network can be detected as a single community. Experimentally, Label Propagation requires a large number of iterations until convergence, thus it can take long running times on large-scale networks. Another algorithm based on the label propagation technique is the Speaker-Listener Label Propagation (SLPA) \cite{xie2011slpa}. SLPA is a linear time algorithm to detect disjoint and overlapping community structure in complex networks, employing a general speaker-listener information propagation process. Experimentally, SLPA present the same variability in the resulting community structure for different runs over the same network. Recently, a fast algorithm was proposed in \cite{han2016massivenets}. It uses a two-step method to build the community structure based on label propagation. In the first step, the structural similarity of nodes is computed for each pair of adjacent nodes. Additionally, it applies the label propagation algorithm to detect meta-communities conformed of most similar nodes. In the second step, a multilevel label propagation technique is applied to build communities that meet a modified version of the Weak community definition \cite{radicchi2004defining}. The second step is based on two sub-steps: network collapse and label propagation. The algorithm introduces a cohesion parameter $\alpha$ (a real value within the range [0, 1]) to control the resolution of the resulting communities. The Community Overlap PRopagation Algorithm (COPRA) \cite{gregory2010finding} is an adaptation of the original Label Programation algorithm to detected overlapping communities. COPRA performs the same propagation algorithm, but allows each node to belong to more than one community. Fluid Communities \cite{pares2017fluid} is a recent algorithm based on label propagation technique, however this algorithm is supervised, because it requires the number of communities as parameter. A Local-First Discovery Method for Overlapping Communities \cite{coscia2012demon} (DEMON) allows nodes to vote for local communities in their ego-neighborhood using a label propagation technique, then the local communities are merged into a global collection of communities.

\textbf{Structural Clustering}: The SCAN \cite{xu2007scan} algorithm and its variants, pSCAN \cite{chang2017mathsf}, SCAN++ \cite{shiokawa2015scan++}, Index-Based SCAN \cite{wen2017efficient} cluster dense zones of nodes determined by the structural similarity of nodes. They are fast but strongly depend on a minimum similarity parameter $\varepsilon$ that is difficult to estimate. To overcome the problem of estimating $\varepsilon$, the parameter-less algorithms SHRINK-H and SHRINK-G \cite{huang2010shrink} were proposed. SHRINK-H performs agglomerative hierarchical clustering by merging dense pairs into micro-communities that conforms the hierarchy of communities, and the final clustering is determined by the partition that maximize the modularity. SHRINK-H presents a worst-case time complexity making it unable to handle efficiently large datasets. SHRINK-G is faster but it sacrifices the capacity of finding hierarchical community structure.

\textbf{Random Walks:} WalkScan \cite{hollocou2016improving}, HeatKernels \cite{kloster2014heat}, PersonalizedPageRank \cite{lofgren2016personalized} and LEMON \cite{li2015uncovering} are recently proposed algorithms to compute local community structure based on the simulation of random-walk processes from seed nodes. The main limitation of those algorithms is the lack of clear methodology to choose the seed nodes, also the number of detected communities strongly depends on the number of selected seed nodes, making them basically supervised algorithms for community detection. Another popular algorithm is the Markov Cluster Algorithm (MCL) \cite{dongen2000cluster}. MCL simulates flow of information in a graph by using a stochastic process composed of two steps: inflation and contraction. The intuition behind MCL is that there is a high flow of information inside dense clusters, but the information rapidly evaporates when it flows between sparsely connected clusters.

\textbf{Others:} Other recently proposed algorithm is Attractor \cite{shao2014community}. Attractor investigates local distance dynamics among connected nodes. Attractor computes the distance on edges based on the Jaccard similarity and applies 3 interaction patterns until the distances converge. The resulting communities are the connected components generated through the removal of the edges with final distance of one. Attractor introduces a cohesion parameter $\alpha$ (a real value within the range [0, 1]) to control the resolution of the detected communities. Greedy Clique Expansion (GCE) \cite{lee2010detecting} is an efficient algorithm to perform overlapping community detection by expanding seed communities in order to optimize a fitness function. Those seed communities are the maximal cliques in the graph that are found automatically by the algorithm.

\subsection{HAMUHI-CODE Algorithm}

The HAMUHI-CODE \cite{castrillo2017fast} is a fast algorithm that finds community structure inspired on an Agglomerative Hierarchical Clustering technique (AHC) composed of three steps. First, the local structural similarity for each edge is computed. Second, each node is set in a separate community, then each community $C$ is merged with its most similar adjacent communities, when $C$ does not meet a community definition passed by parameter. This procedure iterates until all detected communities meet the community definition. Finally, each community $C$ is merged with its most similar adjacent communities, when the $C$’s size is less than a threshold passed by parameter. This procedure iterates until all detected communities achieve the minimum size required.

\subsection{Dynamic Structural Similarity}

The Dynamic Structural Similarity (DSS) \cite{castrillo2018DSS} determines the structural similarity of connected nodes in a graph by dynamically diffusing and capturing information beyond the immediate neighborhood of the analyzed nodes. This approach is based on the following  intuition: \emph{Two nodes are structurally similar if they share an structurally similar neighborhood}. Following the intuitive definition the $DSS(u, v)$ is defined recursively as follows,

\begin{equation}
\label{dss_function}
	DSS(u, v) = \frac{\displaystyle\sum_{\substack{x \in N[u] \cap N[v]}}DSS(u, x) + DSS(v, x)}{\sqrt{\displaystyle\sum_{\substack{x \in N(u)}}DSS(u, x) \times \displaystyle\sum_{\substack{y \in N(v)}}DSS(v, y)}}
\end{equation}

The Equation \ref{dss_function} can be solved by fixed point iteration in super-linear time, so it is able to analyze large-scale graphs.

\section{Improved HAMUHI-CODE}
\label{sec_hamuhi}

\subsection{Disjoint Community Detection}
\label{sub_imp_hamuhi}

The HAMUHI algorithm takes full advantage of the local structural similarity (a.k.a. Cosine similarity) to perform community detection \cite{castrillo2017fast}. However, it has been shown that the local structural similarity presents some limitations that bypass important structural patterns of connections beyond the locality \cite{castrillo2018DSS}. For that reason, we consider that a possibility to improve the quality of the community structure detected by the HAMUHI algorithm, is to support the cluster construction on a robust similarity measure capable of differentiating with higher quality inter-cluster edges from intra-cluster edges. So, an improved version of the algorithm can be achieved after replacing the local structural similarity used in HAMUHI with the DSS, since the DSS increases the probability of identifying intra-cluster edges and inter-cluster edges.

\subsection{Overlapping Community Detection}

In the seminal work developed in \cite{Chakraborty2015LeveragingDC}, it is shown through empirical evaluation, that the overlapping community structure in a network only differs from the disjoint part in the nodes with multiple memberships, i.e., if the overlapping nodes are removed or assigned to a single community, then the filtered community structured corresponds to that detected with classical disjoint community detection algorithms.

Following the previous idea, we propose an extension of the HAMUHI algorithm (O-HAMUHI) to detect overlapping community structure from core disjoint communities. The O-HAMUHI algorithm works with the following two-step approach:

\begin{itemize}

\item \textbf{Step 1}. A disjoint community structure $C(G)$ is detected with the HAMUHI algorithm.

\item \textbf{Step 2}. A fuzzy community structure $C^f(G)$ is generated by computing a membership probability function $f_i:C_i\mapsto[0, 1]$. Optionally, a crisp community structure $C^\alpha(G)$ can be generated from the fuzzy one by applying an $\alpha$-cut to $C^f(G)$.
\end{itemize}

\subsubsection{Membership Probability Function}

In order to define the membership probability function $f_i$, first we define the connectivity of a node $u$ towards a community $C$.

\begin{defn}
The connectivity of a node $u$ towards a community $C$, denoted by $Conn(u, C)$, quantifies the density of connections going from the node $u$ towards nodes in the community $C$ as follows,

\begin{equation}
\label{equ_conn}
 Conn(u, C) = \sum_{v \in C} DSS(u, v)
\end{equation}

\end{defn}

Instead of using other similarity measures to determine the connectivity of a node towards a particular community, like the \emph{Permanence} function proposed in \cite{Chakraborty2015LeveragingDC}, we employ the Dynamical Structural Similarity, since it describes with high quality the density of the neighborhood that surrounds two connected nodes (That is the motivation of the \emph{Permanence} function). This re-utilization saves time of computation, due to the DSS is computed previously in Step 1 of the O-HAMUHI algorithm. Based on the definition of connectivity, the membership probability function is defined as follows,

\begin{defn}
The \emph{Membership Probability} of a node $u$ in a community $C_i$, denoted by $f_i(u)$, is defined as the connectivity $Conn(u, C_i)$ normalized by the maximum connectivity in the neighborhood of $u$, weighted by the fraction of nodes in the community $C_i$ that are connected to the node $u$, that means,

\begin{equation}
\label{equ_membership_probability}
 f_i(u) = \frac{Conn(u, C_i)}{\displaystyle\max_{C_j \in C(G)} Conn(u, C_j)} \times \frac{|\{v \in C_i : (u, v) \in E\}|}{|C_i|}
\end{equation}
\end{defn}

In order to know how representative is $Conn(u, C_i)$, it is normalized respect to the maximum connectivity in the neighborhood of $u$. However, the normalized connectivity can be too high even if the number of connections towards $C$ is low (Due to high density of connections among low number of nodes). For that reason, the normalized connectivity is weighted by the fraction of nodes in $C_i$ that are connected to $u$ (This weighting scheme also gives importance to the number of connections from $u$ towards $C_i$). From Equation \ref{equ_membership_probability} is easy to see that $f_i(u) \in [0, 1]$.

\subsubsection{$\epsilon$-Core Community Definition}

A possibility to increase the probability of identifying overlapping communities, is to detect early in-between communities in the disjoint community structure, for that reason, we propose the following \emph{$\epsilon$-Core Community Definition}.

\begin{defn}
An \emph{$\epsilon$-Core Community} is a candidate community $C$ in Steps 2 and 3 of the HAMUHI algorithm, whose all adjacent communities are connected to $C$ with similarity measure no less than $|maxS-\epsilon|$, where $maxS$ is the maximal similarity measure connecting $C$ to some of its adjacent communities.
\end{defn}

\begin{figure}[t]
\centering
\medskip
\includegraphics[width=60mm]{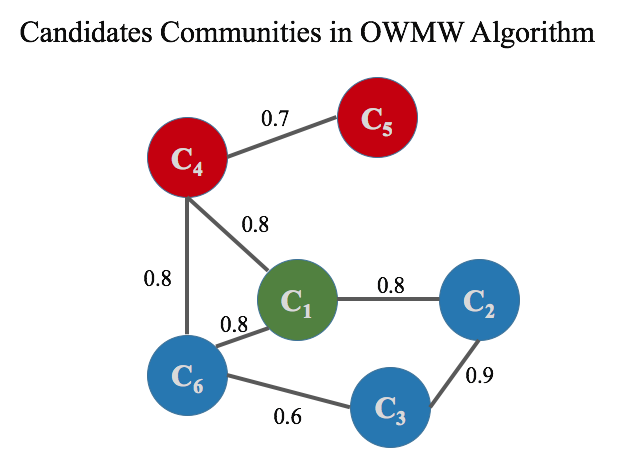}
\medskip
\caption{Six candidate communities in one iteration of the O-HAMUHI algorithm. The community in green color ($C_1$) is an example of an $\epsilon$-Core Community for $\epsilon = 0$. Only the edges with maximum similarity between two communities are displayed.}\label{fig_occ}
\end{figure}

\begin{figure}[!t]
\centering
\medskip
\includegraphics[width=60mm]{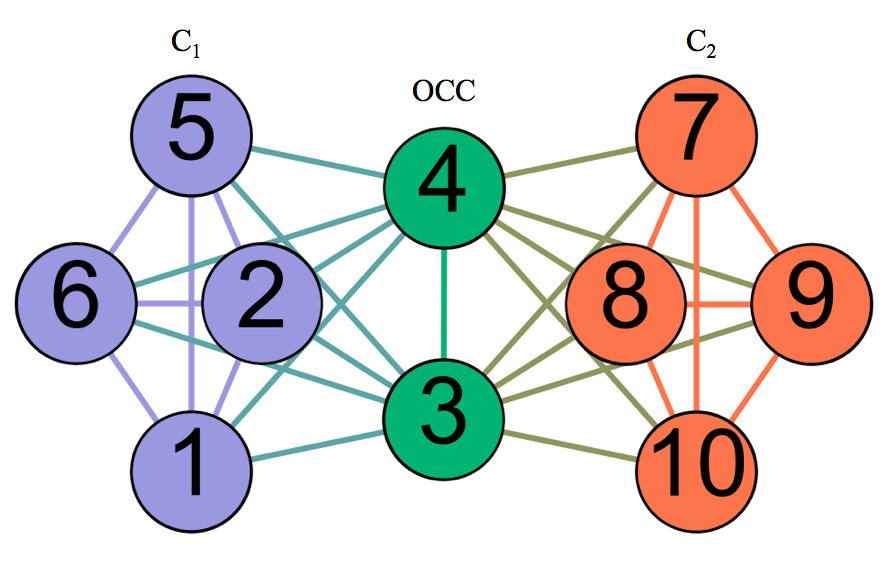}
\medskip
\caption{Toy network with two overlapping communities $C_1$ and $C_2$. The overlapping nodes between $C_1$ and $C_2$ are those in the $\epsilon$-Core  Community $OCC$ composed of the nodes 3 and 4. The algorithm O-HAMUHI correctly assigns the nodes 3 and 4 in both communities $C_1$ and $C_2$ with maximal membership probability.}\label{fig_toy_occ}
\end{figure}

The HAMUHI algorithm must be modified in Step 1 of the O-HAMUHI algorithm in order to detect both, the target community definition passed by parameter and also the $\epsilon$-Core communities. Figure \ref{fig_occ} shows a visual description of an $\epsilon$-Core Community. Additionally, Figure \ref{fig_toy_occ} shows an example of an $\epsilon$-Core conformed by two nodes in a toy network with overlapping community structure. Thanks to the community $OCC$, the O-HAMUHI algorithm correctly classifies the nodes 3 and 4 as the overlapping nodes between the two disjoint communities $C_1$ and $C_2$, and assigns them in both communities with maximal membership probability. Also, by definition, the community $OCC$ is neither a Weak nor a Most Weak community.

\subsection{Complexity Analysis}

The Improved HAMUHI (See \ref{sub_imp_hamuhi}) presents a time complexity of $O(T \times alpha(G) \times |E|)$, such that $\alpha(G) \leq \sqrt{|E|}$ is the arboricity of $G$ \cite{castrillo2017fast}, and $T$ is the the number of iterations performed by the back-end DSS. Because $T$ does not scale with the size of network ($T \approx 5$) \cite{castrillo2018DSS}, it can be considered a constant factor in most practical cases, therefore we argue that this improved version keeps the same asymptotic time complexity of the original HAMUHI algorithm, that is, $O(\alpha(G) \times |E|)$.

For the overlapping extension, we have that the time complexity in Step 1 is dominated by the complexity of the HAMUHI algorithm, thus Step 1 presents  average time complexity of $O(\alpha(G) \times |E|)$. The time complexity in Step 2 is dominated by the time required to compute the membership probability function $f_i$ (Equation \ref{equ_membership_probability}) for each node in the graph. Moreover, to compute the connectivity of a particular node $u$ towards its adjacent communities with Equation \ref{equ_conn}, takes time complexity of $O(d[u])$, since querying the DSS takes $O(1)$ (because the DSS was computed in Step 1). Also, the maximum connectivity in the neighborhood of $u$ and the fractions of nodes connecting $u$ towards a particular community, can be tracked while computing its connectivity. So, to compute the membership probability function for each node in the graph takes time complexity of $\sum_{u \in V} O(d[u]) = O(|E|)$. Therefore, the O-HAMUHI algorithm presents average time complexity of $O(\alpha(G) \times |E|)$, like the original HAMUHI algorithm. The space complexity is still $O(|V|+|E|)$.

\section{Experiments}
\label{sec_experiments}

In this section, several experiments are performed in order to test the properties, efficiency and efficacy of the proposed algorithms and to compare them to the state-of-the-art.  

\subsection{Experimental Settings}
\label{sec_exp_settings}

The experimental settings were chosen from the literature. Those settings are widely used to perform validation, evaluation and comparative analyses among different algorithms, so they provide a solid experimental framework.

\subsubsection{State-of-the-art algorithms}
\label{sub_sota}

In order to compare our proposed algorithms, a set of algorithms was chosen from literature. Those algorithms are top because of their efficiency and efficacy in the community detection task. Only unsupervised algorithms with average time complexity of $O(|E|)$ in complex networks are considered, because they not require the number of communities to detect as parameter, and are applicable to large-scale networks. Table \ref{table_sota} summarizes the selected algorithms.

\begin{table}[t]
\centering
\caption{Set of fast and unsupervised state-of-the-art algorithms selected for the comparative analysis.}\label{table_sota}

\begin{tabular}{ |c|c|c| }
\hline
Algorithm & Detect Overlapping Communities & Reference \\ \hline
Label Propagation (LP) & No & \cite{raghavan2007near} \\ \hline
COPRA & Yes & \cite{gregory2010finding} \\ \hline
SLPA & Yes & \cite{xie2011slpa} \\ \hline
MCL & No & \cite{dongen2000cluster} \\ \hline
ATTRACTOR & No & \cite{shao2014community} \\ \hline
LOUVAIN & No & \cite{blondel2008fast} \\ \hline
SCAN & No & \cite{xu2007scan} \\ \hline
ISCAN & No & \cite{castrillo2018DSS} \\ \hline
pSCAN & No & \cite{chang2017mathsf} \\ \hline
INFOMAP & Yes & \cite{rosvall2008maps} \\ \hline
GCE & Yes & \cite{lee2010detecting} \\ \hline
DEMON & Yes & \cite{coscia2012demon} \\ \hline
\end{tabular}
\centering
\end{table}

\subsubsection{LFR Synthetic Benchmark Graphs}
\label{sub_synt_nets}

The LFR benchmark \cite{lancichinetti2008benchmark} was used to generate the set of synthetic graphs. The LFR benchmark generates unweighted and undirected graphs with planted ground-truth community structure. Also, it produces networks with node degree and community size that follow power-law distributions, making it more appropriate than the Girvan-Newman benchmark to model complex networks. By varying the mixing parameter $\mu$, LFR can generate networks with community structure more or less difficult to identify.

Tables \ref{table_lfr1_nets}, and \ref{table_lfr3_nets} show the parameters used in the LFR Benchmark to generate the sets of synthetic networks. The parameter settings for each dataset were extracted from \cite{yang2016comparative, xie2013overlapping}. However, the community size distribution exponent $\tau_2$ was modified in the dataset LFR1 from -1 to -2.5, in order to model more realistic scenarios and effectively test the resolution limit of the compared algorithms.

\begin{table}[t]
\centering
\caption{LFR1 - Dataset with networks of five different sizes.}\label{table_lfr1_nets}

\begin{tabular}{ |c|c| }
\hline
Parameter & Values \\ \hline
Number of nodes $N$ & 233, 482, 1000, 3583, 8916 \\ \hline
Maximum node degree $maxk$ & 0.1$N$ \\ \hline
Average node degree $avgk$ & 20 \\ \hline
Degree distribution $\tau_1$ & -2 \\ \hline
Maximum community size $maxc$ & 0.1$N$ \\ \hline
Minimum community size $minc$ & Default \\ \hline
Community size distribution $\tau_2$ & -2.5 \\ \hline
Mixing parameter $\mu$ & [0.05, 0.075] with step of 0.05 \\ \hline
Overlapping Nodes $O_n$ & 0 \\ \hline
Overlapping Memberships $O_m$ & 0 \\ \hline
\end{tabular}
\centering
\end{table}

\begin{table}[t]
\centering
\caption{LFR3 - Dataset with networks of size 1000 and 5000 with Big (B) and Small (S) overlapping communities.}\label{table_lfr3_nets}

\begin{tabular}{ |c|c| }
\hline
Parameter & Values \\ \hline
Number of nodes $N$ & 1000, 5000 \\ \hline
Maximum node degree $maxk$ & 50 \\ \hline
Average node degree $avgk$ & 10 \\ \hline
Degree distribution $\tau_1$ & -2 \\ \hline
Maximum community size $maxc$ & 50, 100 \\ \hline
Minimum community size $minc$ & 10, 20 \\ \hline
Community size distribution $\tau_2$ & -1 \\ \hline
Mixing parameter $\mu$ & [0.05, 0.075] with step of 0.05 \\ \hline
Overlapping Nodes $O_n$ & 500 \\ \hline
Overlapping Memberships $O_m$ & [1, 8] with step of 1 \\ \hline
\end{tabular}
\centering
\end{table}

\subsubsection{Validation and Evaluation Criteria}

Many algorithms to perform community detection have been proposed in the literature \cite{fortunato2016community}. For that reason, a standard and reliable framework is precised to validate and evaluate the results obtained with those algorithms and to perform comparative analysis among them.

If networks with ground-truth disjoint communities are provided, a community detection algorithm is validated through the sqrt-Normalized Mutual Information (NMI) \cite{yang2016comparative}. The NMI compares two partitions generated from the same dataset by assigning a score within the range [0, 1], where 0 indicates that the two partitions are independent from each other and 1 if they are equal. 

If the ground-truth communities contain overlapping communtities, the NMI extended to covers \cite{mcdaid2011normalized} is used. If the number of communities in the covers being compared are to different, the NMI is not a good quality measure. For that reason, the Adjusted Omega Index \cite{Collins1988OmegaAG} is also employed. The Adjusted Omega Index assigns a score within the range [0, 1], where 0 indicates that the two partitions are independent from each other and 1 if they are equal. 

\subsection{Disjoint Community Detection}
\label{sub_exp_disjoint}

In order to test the capacity of the Improved HAMUHI algorithm to detect ground-truth communities, it was executed on the 30 LFR networks generated for each combination of parameters listed in Table \ref{table_lfr1_nets}. The mean and standard deviation of the NMI were computed to measure the quality of the results. Moreover, the parameters of HAMUHI were set to minimum community size $K = 2$ and community definition $CD = $MOST WEAK. The default parameter settings were used for the other algorithms.

\begin{figure}[t]
\centering
\includegraphics[width=90mm]{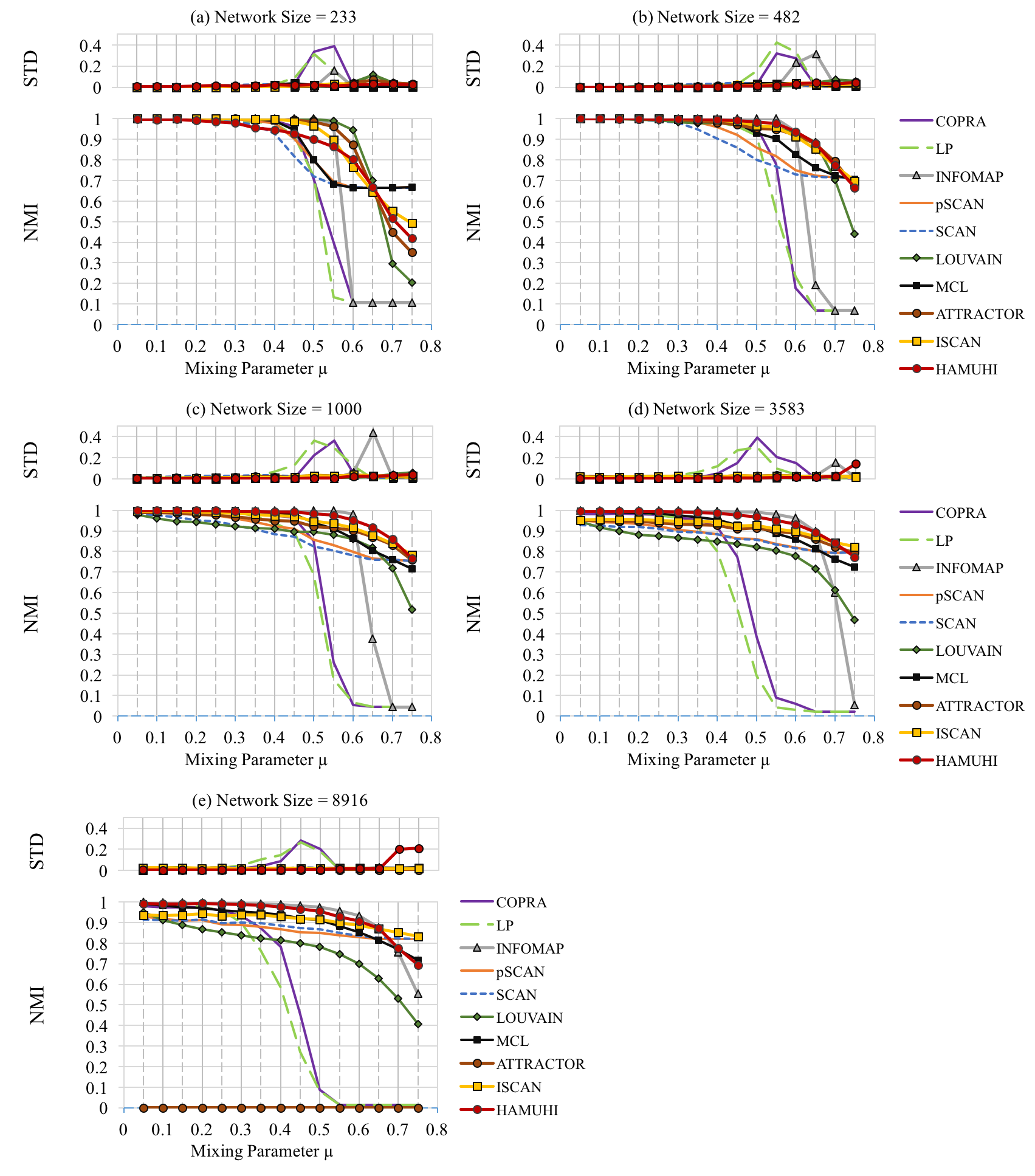}
\caption{(Lower row) Mean value of the sqrt-Normalized Mutual Information NMI (higher values are better) as function of the mixing parameter $\mu$. (Upper row) The standard deviation STD (lower values are better) of the NMI as function of $\mu$.}
\label{fig_nmi}
\end{figure}

As we can see in Figure \ref{fig_nmi}, the critical point on the performance for the majority of the algorithms, arrives when the mixing parameter $\mu > 0.5$. LP and COPRA present good performance until $\mu \approx 0.4$, but after that point they become particularly erratic, presenting high standard deviation. This unstable behavior is due to the random nature of the diffusion process performed by those algorithms. The structural clustering algorithms SCAN, pSCAN and ISCAN offer stable results, but the NMI starts to decrease from low values of $\mu$. However, ISCAN effectively improve the NMI compared to SCAN and pSCAN for each test scenarios, evidencing again the advantages of using the DSS to model dense sub-graphs. Moreover, the LOUVAIN algorithm performance is affected notably when the size of the network increases, due to its resolution limit. INFOMAP is the best performer algorithm until $\mu = 0.6$, offering almost perfect and stable results, but for values of $\mu > 0.6$ it shows an abrupt decay in its performance. In this experiment MCL proof experimentally its efficacy and stability to perform community detection, since it outperforms the majority of algorithms in the test scenarios. ATTRACTOR is a top algorithm since it offers high quality and stable results, however it did not finished the test for the networks of size 8916 (Figure \ref{fig_nmi}c) in a time gap of nine hours. This bad result is due to the convergence problems presented by its underlying dynamic interaction system, even in small networks (a size of 8916 nodes is considered small). 

On the other hand, HAMUHI performs near to the optimal with stable results while $\mu \leq 0.5$, with exception of networks with size 233 where it presents its worst rates of NMI. Moreover, HAMUHI outperforms the other algorithms in the majority of scenarios for values $\mu > 0.6$. HAMUHI has not been tested with the parameter $CD = $WEAK because for values of $\mu > 0.5$ no community in the generated ground truth meets the WEAK definition (making the ground truth undetectable for this parameter setting), in consequence HAMUHI will presents an abrupt change of behavior after the critical point, as evidenced in \cite{castrillo2017fast}. In contrast, HAMUHI presents a smooth transition after the critical point by using the MOST WEAK definition.

\subsection{Overlapping Community Detection}
\label{sub_exp_overlapping}

In order to test the capacity of the O-HAMUHI algorithm to detect ground-truth overlapping communities, it was executed on the 30 LFR networks generated for each combination of parameters listed in Table \ref{table_lfr3_nets}. To get consensus on the results, the mean and standard deviation of the O-NMI and also the Adjusted Omega Index were computed to measure the quality of the results. For this experiment, every algorithm was executed on each test network for different parameter settings, and only the best result obtained among all the parameter settings was averaged in the result. The parameter settings used for each algorithms are the following: For O-HAMUHI the parameters were fixed to $K = 2$, $CD = $MOST WEAK and $T = 5$. Additionally, the crisp threshold was varied for [0.005, 0.01, 0.02, 0.03, 0.04, 0.05]. For SLPA, the crisp threshold was varied for [0.05, 0.1, 0.2, 0.3, 0.4, 0.5]. For COPRA, the maximum number of membership per node was varied in the interval [1, 8]. For DEMON, the crisp threshold was varied for [0.1, 0.15, 0.2, 0.25, 0.3, 0.35]. For GCE the parameter FitnessExponent was varied for [0.8, 0.9, 1.0, 1.1, 1.2, 1.3, 1.4] and the other parameters were set to their default values. INFOMAP was executed with the flag \emph{--overlapping}.

\begin{figure}[t]
\centering
\includegraphics[width=90mm]{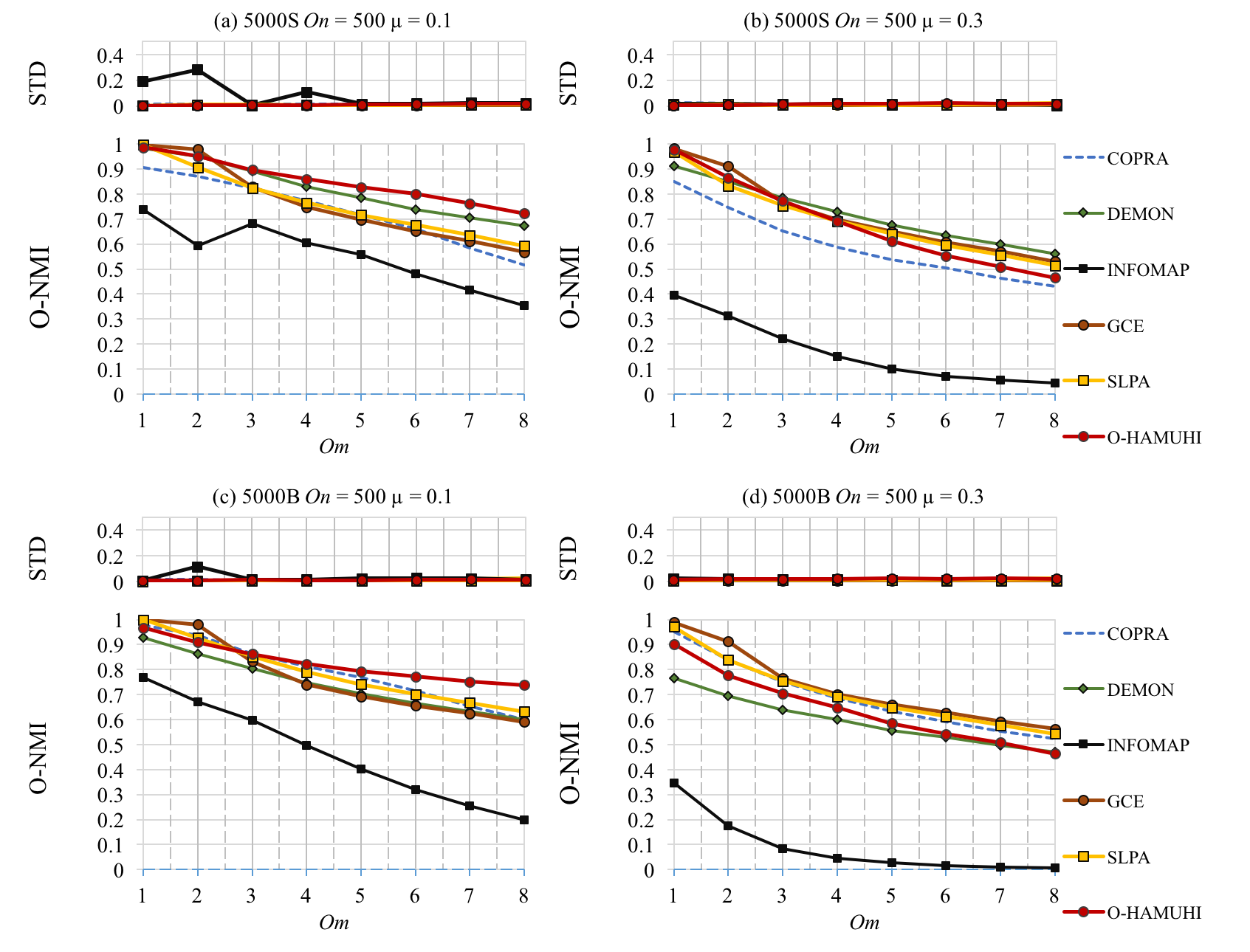}
\caption{(Lower row) Mean value of the Overlapping Normalized Mutual Information, O-NMI (higher values are better) as function of the number of overlapping memberships, $O_m$. (Upper row) The standard deviation, STD (lower values are better) of the O-NMI as function of $O_m$.}
\label{fig_onmi}
\end{figure}

As we can see in Figure \ref{fig_onmi} O-HAMUHI obtains consistently the best averages of O-NMI for values of $O_m \geq 3$ in networks with mixing parameter $\mu = 0.1$. For mixing parameter $\mu = 0.3$, OWMM is outperformed by other algorithms, however O-HAMUHI is still a competitive algorithm. All the algorithms present a similar behavior as $O_m$ increases and their performance is relatively close, with exception of INFOMAP that presents poor results in all the test scenarios.

\begin{figure}[t]
\centering
\includegraphics[width=90mm]{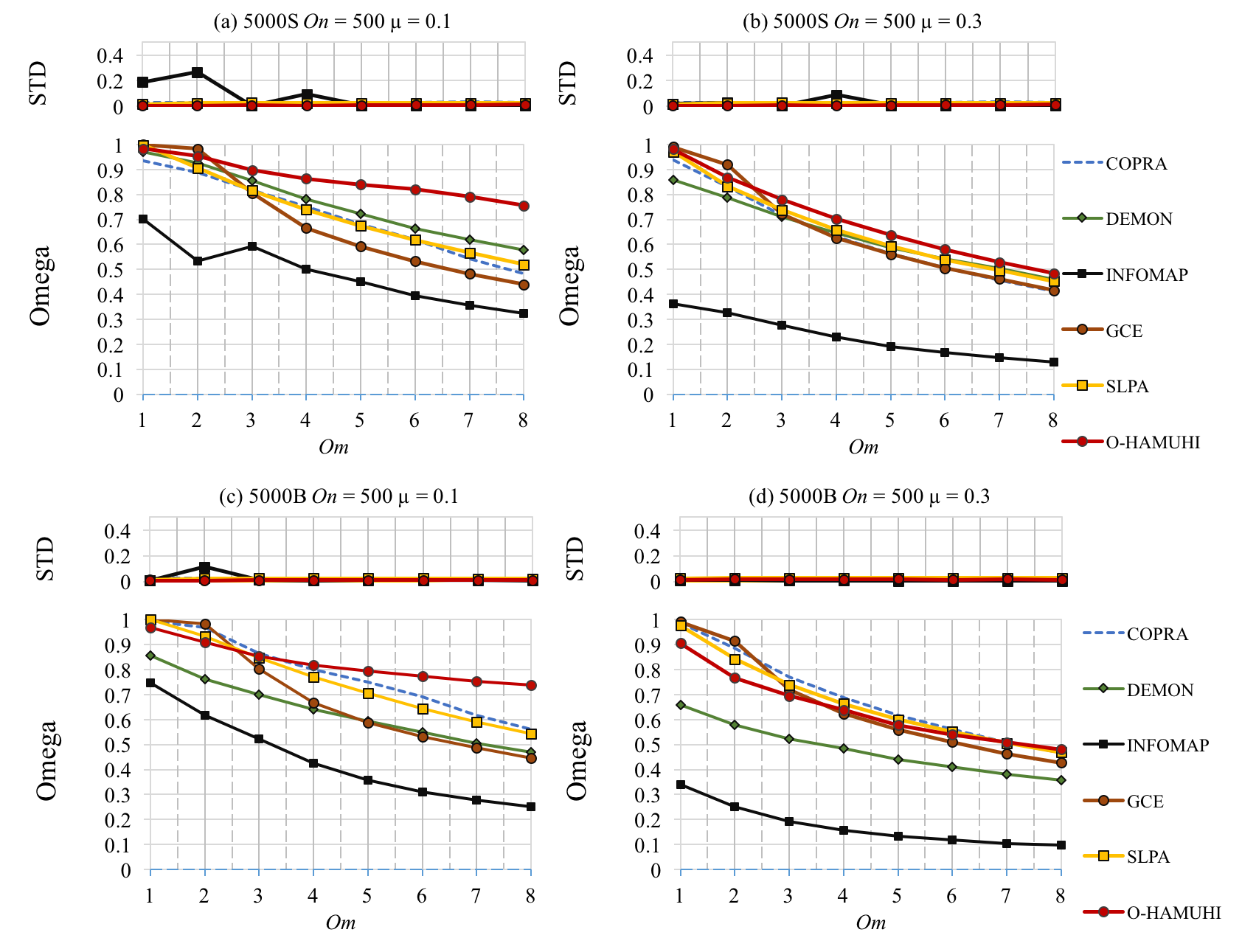}
\caption{(Lower row) Mean value of the Adjusted Omega Index (higher values are better) as function of the number of overlapping memberships $O_m$. (Upper row) The standard deviation, STD (lower values are better) of the Adjusted Omega Index as function of $O_m$.}
\label{fig_omega}
\end{figure}

A similar tendency can be observed in Figure \ref{fig_omega}. In this case O-HAMUHI presents the best average performance in terms of the Omega Index for values of $O_m \geq 3$ in all scenarios, with exception of \ref{fig_omega}d where it is still competitive. Moreover, in networks with mixing parameter $\mu = 0.1$ and for values of $O_m \geq 5$, the performance of O-HAMUHI deviates significantly from the others by achieving an improvement of up-to 20 percent in the best case.

In all the test scenarios, the majority of algorithms provide stable results, with standard deviation below $0.03$.

\section{Conclusions}

In this paper we proposed an improved version of the  HAMUHI algorithm. Thanks to the back-end Dynamic Structural Similarity, the improved version can increase its efficacy in the disjoint community detection task. Also, we proposed an extension of the previous algorithm that leverages the disjoint community structure to efficiently detect fuzzy and crisp overlapping community structure. Three core elements have been proposed to compute the overlapping community structure: \emph{i)} The connectivity function that quantifies the density of connections of a node towards a particular disjoint community, relying its computation on the Dynamic Structural Similarity measure. \emph{ii)} The $\epsilon$-Core community definition that increases the probability of identifying in-between communities in the disjoint community structure. \emph{iii)} The membership probability function to compute the soft partition from the core disjoint communities.

The experimental evaluation shows that our proposals can detect efficiently high quality disjoint and overlapping community structure in complex networks, and compared to several top state-of-the-art algorithms, our proposals obtain superior results in several scenarios, so they can be considered good candidates to perform community detection. Moreover, the improved and extended algorithms keep the same computational complexity of the original HAMUHI algorithm, thus it is still applicable to large-scale complex networks. 

As future work we propose to adapt the proposed algorithms to dynamic networks generated from data streams.

\label{sec_conclusions}


\end{document}